\documentclass[sigconf]{acmart}

\AtBeginDocument{%
  }

\copyrightyear{2018}
\acmYear{2018}
\acmDOI{XXXXXXX.XXXXXXX}

\acmConference[Conference acronym 'XX]{Make sure to enter the correct
  conference title from your rights confirmation emai}{June 03--05,
  2018}{Woodstock, NY}
\acmISBN{978-1-4503-XXXX-X/18/06}

\usepackage{multirow}    
\begin{document}
\newcommand{\Quote}[1]{
\begin{quote}
   \textit{#1} 
\end{quote}
}

\title{Futuring Social Assemblages: How Enmeshing AIs into Social Life Challenges the Individual and the Interpersonal}

\author{Lingqing Wang}
\affiliation{%
  \institution{Georgia Institute of Technology}
  \city{Atlanta}
  \country{USA}}
\email{lwang809@gatech.edu}

\author{Yingting Gao}
\affiliation{%
  \institution{Georgia Institute of Technology}
  \city{Atlanta}
  \country{USA}}
\email{ygao617@gatech.edu}

\author{Chidimma Lois Anyi}
\affiliation{%
  \institution{Georgia Institute of Technology}
  \city{Atlanta}
  \country{USA}}
\email{canyi3@gatech.edu}

\author{Ashok Goel}
\affiliation{%
  \institution{Georgia Institute of Technology}
  \city{Atlanta}
  \country{USA}}
\email{ashok.goel@cc.gatech.edu}

\renewcommand{\shortauthors}{Wang et al.}

\begin{abstract}
Recent advances in AI are integrating AI into the fabric of human social life, creating transformative, co-shaping relationships between humans and AI. This trend makes it urgent to investigate how these systems, in turn, shape their users. We conducted a three-phase design study with 24 participants to explore this dynamic. Our findings reveal critical tensions: (1) social AI often exacerbates the very interpersonal problems it is designed to mitigate; (2) it introduces nuanced privacy harms for secondary users inadvertently involved in AI-mediated social interactions; and (3) it can threaten the primary user's personal agency and identity. We argue these tensions expose a problematic tendency in the user-centered paradigm, which often prioritizes immediate user experience at the expense of core human values like interpersonal ethics and self-efficacy. We call for a paradigm shift toward a more provocative and relational design perspective that foregrounds long-term social and personal consequences.
\end{abstract}



\maketitle

\section{Introduction}
\begin{quote}
    “We make our technologies, and they, in turn, shape us. So, of every technology we must ask, Does it serve our human purposes?” 
    \newline — Sherry Turkle, \textit{Alone Together: Why We Expect More From Technology and Less From Each Other} \cite{10.5555/1972496}
\end{quote}
The rapid integration of AI into most aspects of human life, from birth \cite{viswanathan_interaction_2025, lee_application_2020} to death \cite{brubaker_ai_2024, morris_generative_2025}, from personal reflection \cite{g_mitchell_reflection_2021, kocielnik_designing_2018, kocielnik_reflection_2018, lo_autoethnographic_2024} to juridical decision-making \cite{kolkman_justitia_2024, grgic-hlaca_human_2019}, demands critical examination. The symbiotic relationships between humans and the things they create raise cautions about our design choices of AI systems. The stakes are particularly high, as technology has long been constitutive of humanity, i.e., we are not who we are today without the technology we use \cite{verbeek_cyborg_2008, verbeek_beyond_2015}. In other words, by designing the AI for our lives, we are actively designing future user-AI assemblages, where AIs and users form newfound entities, and they iteratively co-adapt, transforming what it means to be human.

Such assemblages' evolution is gradually revealed in many fields, presenting challenges and opportunities for future design \cite{kumar_mobile_2013, abbass_social_2019, kleinberg_algorithmic_2021, carli_human-agent_2024}. Studies have raised concerns about human autonomy \cite{kumar_mobile_2013, abbass_social_2019}, homogenization \cite{kleinberg_algorithmic_2021}, value shifts \cite{bernstein_embedding_2023}, and deskilling in human-AI assemblages at both personal and societal levels \cite{carli_human-agent_2024}. Empirical studies have predominantly explored the effect of such assemblages in task-oriented scenarios with well-defined metrics, such as efficiency \cite{gebreegziabher_supporting_2025, ramos2020interactive}, and working productivity and quality \cite{dell2023navigating}. Far less understood are the dynamics of assemblages in social scenarios, where fundamentally different forms of intelligence are required \cite{herrmann_humans_2007, premack_does_1978}. Unlike other tasks, social interactions are more relational and subject to an individual’s lived experience, and deeply embedded in a specific sociocultural context \cite{goffman_presentation_2023, goffman_behavior_2008, goffman_interaction_2017, mead_mind_1934, brown_politeness_1987}. Consequently, it is difficult to directly translate findings from task-based research to the social domain, or to evaluate these less objective and ill-defined social constructs. 
Within social scenarios, research has long been designing AI functions for human social life, from augmenting interpersonal messages \cite{fu_comparing_2023, hohenstein_ai-supported_2018, hancock_impression_2001, hancock_ai-mediated_2020, hohenstein_artificial_2023} to simulating loved ones \cite{xygkou_conversation_2023, morris_generative_2025, brubaker_ai_2024}. Yet, much of this work prioritizes system capabilities or immediate interactional effects, often rendering the sociocultural infrastructure opaque \cite{10.1145/1054972.1054998}. Given the increasing normalization of AI in social scenarios \cite{phang2025investigating}, there is an urgent need to look beyond immediate utility and ask: how might the co-adaptive integration with social AI transform users and human sociality? 

To envision the transformative effects, we present a three-step design study inspired by previous works \cite{petsolari_socio-technical_2024, showkat_its_2022, wong2023broadening}. We created future-oriented design fictions to explore the possibilities and implications of incorporating AI technologies into the context of digital social interaction. First, we engaged with 12 participants to explore the design space of AI support for digital social interaction. Second, grounded in participants’ perspectives and the literature, we created fifteen design fictions that situate social AI support in the digital social space to foster conversation about supporting future digital social interactions. Finally, we utilized the created storyboards to elicit participants' visions of future social AI through 18 semi-structured interviews. 

Our research reveals that at the interpersonal level, participants are concerned that the integration of AI into social life will erode interpersonal connections and authenticity, paradoxically deepening the problems of social isolation that such technologies often claim to solve. For non-primary\footnote{We distinguish primary users and non-primary users similarly as previous work \cite{wong_broadening_2023, h_tan_monitoring_2022}: primary users have greater forms of control and access to social AI, often by owning and operating them, while non-primary users have less awareness, control and access to social AI. For example, the primary user uses an AI to generate replies to emails, while the non-primary user receives these messages without knowing they are interacting with an AI.} users who are inadvertently involved in AI-mediated interpersonal social interactions but have less control and access to the AI, the use of AI in a social context introduces nuanced privacy and consent concerns. For the primary user, this entanglement with AI poses a threat to their social agency and identity, which are fundamentally formed through interaction with others. Furthermore, while participants acknowledged the benefits of quantification by AI, they viewed the attempt to align human complexity with these metrics as a negative form of reductionism.
We argue these challenges are rooted in the dominant user-centered design paradigm, which prioritizes an individual’s immediate experience and satisfaction. Therefore, we offer design implications for shifting toward an interpersonal and provocative design paradigm. This alternative approach would consider relational dynamics, account for non-primary users, and challenge users to engage with social AI in a more ethical and prudent manner.

This paper contributes to Human–AI Interaction by examining how AI might become deeply integrated into people’s social lives, an inherently subjective and nuanced domain. Using speculative methods, we reveal how designs that prioritize short-term convenience in social AI may produce unintended long-term consequences. We offer a contextual understanding of shifting agency and identity within user–AI social assemblages and highlight the complex interpersonal effects of social AI that extend beyond the notion of a single, bounded “user.” Finally, we propose design implications for expanding current user-centered paradigms, advocating a shift toward more relational and provocative approaches that account for broader social dynamics.

\section{Related Work}
\subsection{Conceptualizing Human-AI Co-Adaptation as Entangled Assemblages}
Scholarship on human-AI integration increasingly moves beyond simplistic models of tool use to explore complex, symbiotic relationships. Initial research often followed one of two paths: AI augmenting human processes by leveraging the distinct capabilities of each partner \cite{lyytinen_metahuman_2021, riedl2025ai, schecter_how_2025, mahmud_study_2023}, or an AI-centric view that incorporates humans into the AI lifecycle to improve model performance and alignment, as seen in human-in-the-loop systems \cite{cranshaw_calendarhelp_2017,van_rijn_giving_2024}.
More recent work synthesizes these paths into a bidirectional, co-adaptive process where humans and AI iteratively shape one another \cite{mosqueira-rey_human---loop_2023, ramos2020interactive, peeters2021hybrid}. For example, in interactive machine learning, users teach AI models to label data, thereby refining their own understanding \cite{gebreegziabher_supporting_2025}, while the machine, in turn, guides the human to be a better teacher \cite{ramos2020interactive}. The iterative and co-adaptive interaction between human and AI alludes to an entangled relation where a human and their AI mutually define themselves and co-evolve over time via their situated intra-action, which not just extends human capabilities or opens up the possibilities but shapes us into different people \cite{frauenberger_entanglement_2020, verbeek_beyond_2015, verbeek_cyborg_2008}.

The concept of the ``assemblage'' provides a powerful theoretical lens for understanding this entanglement. An assemblage perspective posits that artifacts are not isolated entities but are in a constant state of reformulation with their environment and users. From a cognitive perspective, \citet{hayles2016cognitive} argues that human-technology assemblages reconfigure the contexts under which human cognition operates. For instance, using digital assistants for navigation may weaken innate human spatial reasoning abilities \cite{hayles2016cognitive}. This reconfiguration occurs across the full cognitive spectrum, including ``consciousness, the unconscious, the cognitive nonconscious, and the sensory/perceptual systems,'' ultimately reshaping what constitutes human intelligence in a technologically saturated society  \cite{hayles2016cognitive}. From a material perspective, the concept of "digital data assemblages" highlights the inextricable link between humans and their data, which constructs dynamic and distributed forms of selfhood \cite{lupton_feeling_2017}.

The formation of these human-AI assemblages carries profound, often ambivalent, societal implications \cite{andres_understanding_2024, abbass_social_2019}. 
The societal integration of AI envisions a future of continuous collaboration, with humans actively participating in decision-making loops alongside AI \cite{abbass_social_2019}. The effectiveness of this human-AI relationship is theorized to be a system of reciprocal influence, critically dependent on factors like trust and strategic task allocation \cite{abbass_social_2019}. 
Such reciprocity is not always beneficial and can create significant challenges. Social media algorithms, for instance, both embed human values and actively shape user perspectives and behaviors \cite{bernstein_embedding_2023}. This can be leveraged for surveillance and control, as algorithmic categorization of users creates new forms of identity that serve commercial or political interests \cite{cheney-lippold_new_2011}. Furthermore, individuals' over-reliance on optimal algorithmic suggestions risks creating an intellectual "monoculture," where the diversity of decision-making across a population is dangerously reduced \cite{kleinberg_algorithmic_2021}. This leads to a broader concern that widespread dependency on AI for fundamental needs could erode the collective intelligence of the human population, even as it enhances individual capabilities \cite{kumar2023humans}.

This body of literature establishes that the relationship between humans and AI is not one of a user and a tool, but of a deeply entangled, co-adaptive assemblage with transformative potential. Building on these conceptualizations, this work investigates the mechanisms of co-adaptation and transformation within human-AI assemblages, with a specific focus on their impact on human social interaction.

\subsection{AI for Social Interaction in Digital Spaces}
Digital social spaces have become fundamental to contemporary life, supporting everything from intimate relationship formation \cite{zytko2023online} and social support networks \cite{ryan2021appropriation, tixier_counting_2016, lei2024unpacking} to professional development \cite{dillahunt_village_2022, park2024who2chat} and collective learning \cite{celina_sole_2016, lee_design_2023}. However, these environments present unique challenges, such as reduced visibility of social cues, limited awareness of potential connections, and decreased social accountability \cite{wang_understanding_2022, zhu_zoombatogether_2023}. To address these challenges, researchers have explored how AI can support and shape human social interaction in three primary capacities: as a mediator, a social agent, and a community facilitator \cite{freeman2025new, lim_designing_2023}.

A primary area of research, AI-Mediated Communication (AI-MC), investigates how AI can operate on behalf of communicators by modifying, augmenting, or even generating messages \cite{hancock_ai-mediated_2020}. Applications that compose or rephrase content for emails and text messages, for example, can significantly alter a user's language \cite{fu_comparing_2023, hohenstein_ai-supported_2018}. This intervention can influence interpersonal perceptions and dynamics, raising important questions about authenticity, ethics, and culture \cite{hancock_ai-mediated_2020, hohenstein_artificial_2023}. AI-MC thus serves to both streamline communication and introduce a layer of algorithmic influence on how we present ourselves and understand others.

Beyond mediation, AI can also function as an autonomous social agent within digital environments. These agents engage in social conversations \cite{hwang_whose_2024, ringfort-felner_kiro_2022, clark_what_2019}, and provide social support to users \cite{xygkou_conversation_2023, piccolo2021chatbots}. The recent rise of generative AI has amplified this potential, leading to explorations of AI that can realistically simulate human behaviors in social interaction \cite{park_social_2022} or even assist human counselors in preparing for online mental health support sessions \cite{park_social_2022}.

At a broader community level, AI offers tools for managing and facilitating group interactions, though these tools introduce a set of societal challenges \cite{lim_designing_2023}. On one hand, AI can foster positive online collaboration by matchmaking individuals for professional or social purposes \cite{zhu_value-sensitive_2018, 10.1145/3544548.3580930, wang_understanding_2022, kakar2024sami}, inducing solidarity between groups and reducing bias  \cite{kong_working_2025, claggett_relational_2025}, and powering community moderation based on shared norms \cite{kou_mediating_2020}. On the other hand, the same algorithmic systems can lead to intellectual isolation and social polarization \cite{rodilosso_filter_2024}. Similarly, while AI avatars in dating scenarios may enhance user engagement, they raise critical concerns about authenticity and the gap between a digital persona and a real-life individual \cite{baradari_data-driven_2025}.

This body of work demonstrates numerous roles AI can play in humans' digital social life. However, much of this research examines either the system's capabilities or its immediate effects on social interactions. Building on these rich threads, our work focuses on the subjective user experience that arises from a deeper integration of AI into one's social life. We specifically explore how the user-AI assemblage is perceived by and interacts with others.

\subsection{Envisioning Human-AI Assemblage}
\label{lit_envisioning}
The emergence of AI as a design material has profoundly expanded the possibilities for human-computer interaction \cite{holmquist_intelligence_2017, yang2018machine, dove_ux_2017, yang_sketching_2019, yang_re-examining_2020, subramonyam_solving_2022}. However, this expansion introduces significant socio-technical challenges, such as accounting for the opaque nature of AI systems \cite{holmquist_intelligence_2017}, integrating nuanced social interpretation \cite{baumer2017toward}, and anticipating broader societal impacts \cite{keyes2019mulching}. Simply designing for functional goals is insufficient; we must also consider how these technologies integrate into and reshape the complex fabric of human relationships and social norms \cite{star1994steps, lee2017bridge}. This requires methods that look beyond immediate use to explore potential futures.

To anticipate the potential impact of AI technologies beyond their immediate goals and outcomes, speculative design approaches are often utilized \cite{showkat_its_2022, ringfort-felner_kiro_2022, luria_letters_2022, petsolari_socio-technical_2024, im2023ai, kiskola2025generative}. 
This set of approaches, including design fiction \cite{baumer2020evaluating, bleecker2022design}, speculative design \cite{dunne_speculative_2013}, and critical design \cite{10.1145/2598510.2598588, bardzell_what_2013}, are commonly used in HCI to imagine alternative presents and possible futures and provoke critical reflection on the potential implications of new technologies \cite{ringfort-felner_quality_2025}. 
For example, researchers have used fictional websites for in-vehicle voice assistants to explore the nuances of fully-fledged human-AI social conversations in the future \cite{ringfort-felner_kiro_2022}, and created fictional scenarios of AI-assisted parenting to gauge the desirability of such futures with parents themselves \cite{petsolari_socio-technical_2024}. By creating tangible, albeit fictional, artifacts, speculative design makes abstract possibilities concrete and debatable.

While powerful for opening up new possibilities, speculative approaches can risk becoming disconnected from the lived realities of the people they are meant to serve. The socio-technical perspective reminds us that technology is never neutral; it co-evolves with users, their practices, and their values over time \cite{star1994steps, lee2017bridge}. And AI tools will inevitably interact with existing artifacts, multiple stakeholders, and evolving norms in the context in which they are deployed.
This has led to a strong emphasis on participatory design and co-design methods in Human-AI interaction \cite{zhang2023stakeholder, lin2021engaging,zytko_participatory_2022, delgado2023participatory, lee_webuildai_2019},  while value-sensitive AI design emphasizes incorporating diverse human values and tacit knowledge directly into the design process, ensuring that systems are grounded in genuine human needs from the earliest stages \cite{zhu_value-sensitive_2018, showkat_its_2022}.

Motivated by this, our study synthesizes these two traditions. We utilize speculative design prompts not as an end in themselves, but as a tool within a participatory framework. This approach enables us to explore the future of human-AI social assemblages while ensuring that the exploration is grounded in the perspectives, values, and experiences of our participants.

\section{Methodology}
\begin{figure*}
    \centering
    \includegraphics[width=1\linewidth]{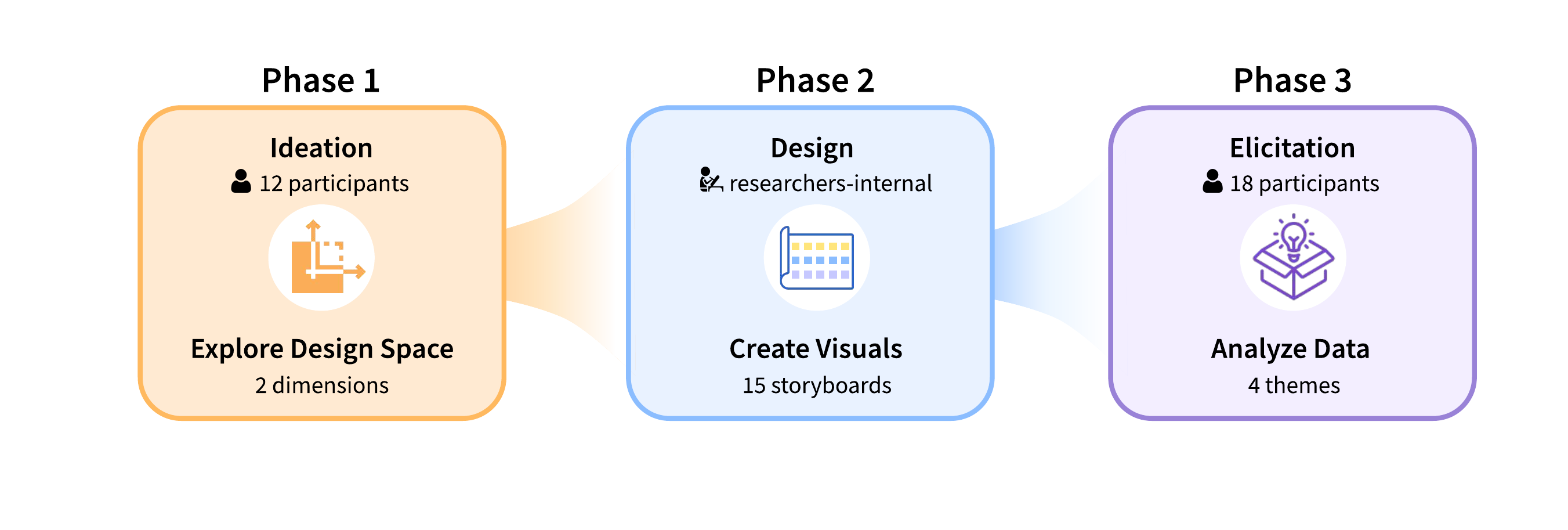}
    \caption{Overview of the three-phrase study}
    \label{fig:method}
\end{figure*}

Our approach is situated at the intersection of participatory and speculative design (Sec. \ref{lit_envisioning}). We are particularly inspired by the methods of several foundational studies in this area \cite{petsolari_socio-technical_2024, showkat_its_2022, wong2023broadening, seering_beyond_2019}. We engaged with 24 participants in total (Table \ref{tab:participants_data}). Through three phases of study (Fig. \ref{fig:method}), we explored how online graduate students envision AI's support and impact on their social lives.
In Phase 1, ideation workshops were conducted with participants to understand their experiences with digital social interaction, develop shared goals, and engage them in exploring the design space. 
In Phase 2, we researchers created fifteen storyboards grounded in the design space identified in Phase 1. 
In Phase 3, elicitation sessions were conducted with participants, utilizing the storyboards from Phase 2 to engage them in reflecting on and discussing future human-AI social assemblages. 

\subsection{Study Scope and Participants}
\label{method_scope}
We recruited participants from a fully online degree program offered to students globally by a higher institution in North America. The recruitment messages were distributed with screening forms through various channels, including program discussion forums, social media, mailing lists, and word-of-mouth. Participants were 18 years of age or older and expressed interest in expanding their social connections online. They were compensated with \$25 per hour for their participation. Twelve participants were recruited for the Phase 1 study. Six of them, plus twelve additional participants, joined the Phase 3 study. The detailed demographic information of our participants is presented in the Table \ref{tab:participants_data}. The study received institutional IRB approval.

We focused our study on forming social connections in an online education program as an example of social interaction. We engaged online graduate students in a computer science program who often face social isolation and challenges in balancing work, life, and academic commitments.  
We chose to focus on this specific group for three primary reasons:
(1) Focusing the research scope on the broad spectrum of social interaction. The application of AI in supporting social interaction is a broad spectrum, ranging from fostering intimate connections \cite{chen_closer_2023} to moderating online forums for safety \cite{kou_mediating_2020}. By situating our research within the life stage of our participants, we narrow our focus to a manageable and relevant segment of the social interaction spectrum. Specifically, this study focuses on AI's role in supporting online peer interactions within educational and professional contexts. Participants also discussed other social connections, reflecting the fluid nature of relationships where colleagues may become friends.
(2) The increasing significance of digital social interaction. Social interaction in the digital sphere is critical for education and professional development \cite{dillahunt_village_2022, park2024who2chat, goel2020ai}, and its prevalence in daily life has grown significantly, particularly since the COVID-19 pandemic.
However, this shift has also exacerbated issues such as social isolation, anxiety, and information overload \cite{woodruff_how_2024, toscano_social_2020, galanti_work_2021, ali2015comparing, leal_filho_impacts_2021}.
This context presents an opportunity for technological interventions designed to facilitate more meaningful and supportive online engagement. As mentioned above, our participants also expressed interest in increasing their digital social interactions.
(3) The participants' technical background benefits the envisioning of AI. We acknowledge that our participant pool, primarily composed of individuals with technical backgrounds, is not representative of the general population, which impacts the generalizability of our findings. However, given the exploratory nature of this study, their familiarity with AI provides a distinct advantage, namely, the conceptual understanding required for nuanced speculation and envisioning future AI systems. This technical grounding enabled a depth of feedback on potential AI roles and features that would be more difficult to elicit from a lay audience, making their contributions particularly valuable for formative design research.

\begin{table*}[]
    \centering
    \caption{Demographics of participants.}
    \begin{tabular}{c c c c c c c} 
    \hline
        ID & \textbf{Age} & \textbf{Gender} & \textbf{Current Location} & \textbf{Profession} & \multicolumn{2}{c}{\textbf{Participation}$^{1}$} \\
        \hline
        P01 & 29 & Male & United States & Energy Performance Engineer & ID & EL \\
        P02 & 26 & Male & Canada & Software Developer & ID &   \\
        P03 & 23 & Male & India & Data Scientist & ID & EL \\
        P04 & 30 & Non-binary & United States & Software Engineer & ID &   \\
        P05 & 24 & Male & India & Software Engineer & ID &   \\
        P06 & 27 & Male & United Arab Emirates & Technical Engineer & ID &   \\
        P07 & 32 & Male & China & Teacher & ID &EL \\
        P08 & 38 & Female & United States & Unemployed & ID &EL \\
        P09 & 32 & Female & United States & Grad Student & ID &   \\
        P10 & 32 & Male & India & Software Engineer & ID &   \\
        P11 & 44 & Male & Sweden & CEO/CTO & ID &EL \\
        P12 & 39 & Female & United States & Data Scientist & ID &EL \\
        P13 & 23 & Male & United States & Academic/Musical Tutor &   &EL \\
        P14 & 41 & Male & United States & Data Engineer &   &EL \\
        P15 & 26 & Male & United States & Programmer &   &EL \\
        P16 & 26 & Male & United States & Software Engineer &   &EL \\
        P17 & 28 & Female & United States & Student &   &EL \\
        P18 & 23 & Male & United States & Data Analyst &   &EL \\
        P19 & 36 & Male & United Kingdom & Academic Researcher &   &EL \\
        P20 & 24 & Male & India & Entrepreneur &   &EL \\
        P21 & 26 & Male & United States & Student &   &EL \\
        P22 & 26 & Female & Canada & Software Engineer &   &EL \\
        P23 & 26 & Male & Mexico & Software Engineer &   &EL \\
        P24 & 28 & Female & United States & Software Engineer &   &EL \\
        \hline
    \end{tabular}
    \caption*{\footnotesize $^{1}$ $ID$: Participation in Phase 1 - Ideation; $EL$: Participation in Phase 3 - Elicitation.}
    \label{tab:participants_data}
\end{table*}

\subsection{Phase 1: Ideation}

To understand how we might design for this specific space, we conducted semi-structured interviews with design tasks to explore participants' experiences, motivations, expectations, and current barriers to forming social connections online, and invited them to imagine ideal social interaction scenarios involving AI and elaborate on specific AI design concepts. We conducted this co-design study with 12 participants. All researchers met regularly, shared each other’s analysis, and engaged in discussion. Similar to previous work \cite{petsolari_socio-technical_2024, seering_beyond_2019, schulte2016homes}, we further grounded the design space by conducting a non-exhaustive literature search of work in communication theory (e.g., \cite{knapp_interpersonal_2014, altman1973social, berger_explorations_1975}), sociology and social psychology of interpersonal interaction (e.g., \cite{goffman_presentation_2023, goffman_behavior_2008, goffman_interaction_2017}), and AI design for social interaction (e.g., \cite{hastings_lift_2020, dagan_synergistic_2021, welge_better_2016}).

A 2-dimensional design space is identified from this process, synthesizing users' ideation from the interview and literature research (see Table \ref{tab:human_ai_lifecycle} and Table \ref{tab:ai_social_capability} for a snapshot of the two dimensions, and Appendix \ref{app:design} for a more comprehensive review of this design space with participants' ideas, with quotes and theoretical support from the literature search).
One dimension is a cyclical process for human-AI assemblages to form online social interaction (Table \ref{tab:human_ai_lifecycle}), which represents the developmental progression of human relationships as depicted in communication theories (e.g., \cite{knapp_interpersonal_2014, altman1973social, berger_explorations_1975}). Across this progress, five key stages of human-AI collaboration were identified from participants' envisioning. The second dimension concerns people's expectation of AI's social capabilities levels (Table \ref{tab:ai_social_capability}), which also aligns with the conceptualizations of AI in earlier research (e.g., \cite{kim2024much, kumar2023humans, sadeghian2021limitations}).

\begin{table*}[h!]
\caption{The Cyclical Process of Forming Social Interaction \textbf{(Dimension 1)}.}
\begin{tabular}{@{} l p{0.75\textwidth} @{}}
\toprule
\textbf{Stage} & \textbf{Human-AI Actions} \\
\midrule
\textbf{S1. Data Collection} & The user grants permissions and sets the scope with the AI to compile personal data. \\
\addlinespace 
\textbf{S2. Profiling} & The AI models user attributes through interaction, where the user actively shapes by imbuing it with personal knowledge. \\
\addlinespace
\textbf{S3. Social Formation} & AI facilitates the user decides how to pursue social connection and to initiate social action. \\
\addlinespace
\textbf{S4. Social Maintenance} & The AI facilitates the user to sustain social momentum, such as actively communicating and engaging with their connections. \\
\addlinespace
\textbf{S5. Reflection} & The user evaluates their social experiences and provides feedback to the AI, which the AI takes into consideration for outcome analysis and refinement. \\
\bottomrule
\end{tabular}
\label{tab:human_ai_lifecycle}
\end{table*}

\begin{table*}[h!]
\caption{AI Social Capability Levels \textbf{(Dimension 2)}.}
\centering
\begin{tabular}{l p{0.7\textwidth}}
\hline
\textbf{Level} & \textbf{Core Concept}\\
\hline
\textbf{L1. Human Augmentation} & Replicates human abilities for routine tasks with greater computational capacity. \\

\textbf{L2. Social Replica} & Incorporates human social nuances to be socially aware and cognizant. \\

\textbf{L3. Omniscient Social Beings} & Synthesizes human knowledge and goes beyond it. \\
\hline
\end{tabular}
\label{tab:ai_social_capability}
\end{table*}

\subsection{Phase 2: Design}
We used the design space identified in Phase 1 as a conceptual model to facilitate design thinking for the envisioning of AI technologies that might support social interaction, inspired by previous studies \cite{petsolari_socio-technical_2024, wong_broadening_2023}. The hypothetical storyboards were designed to (1) conceptually explore the functionality of AI to support social connection grounded in the design space discovered in the first stage; (2) facilitate participants to engage, envision, and reflect the implications of such future happening in their own lives. 

\begin{figure*}
    \centering
    \includegraphics[width=1\linewidth]{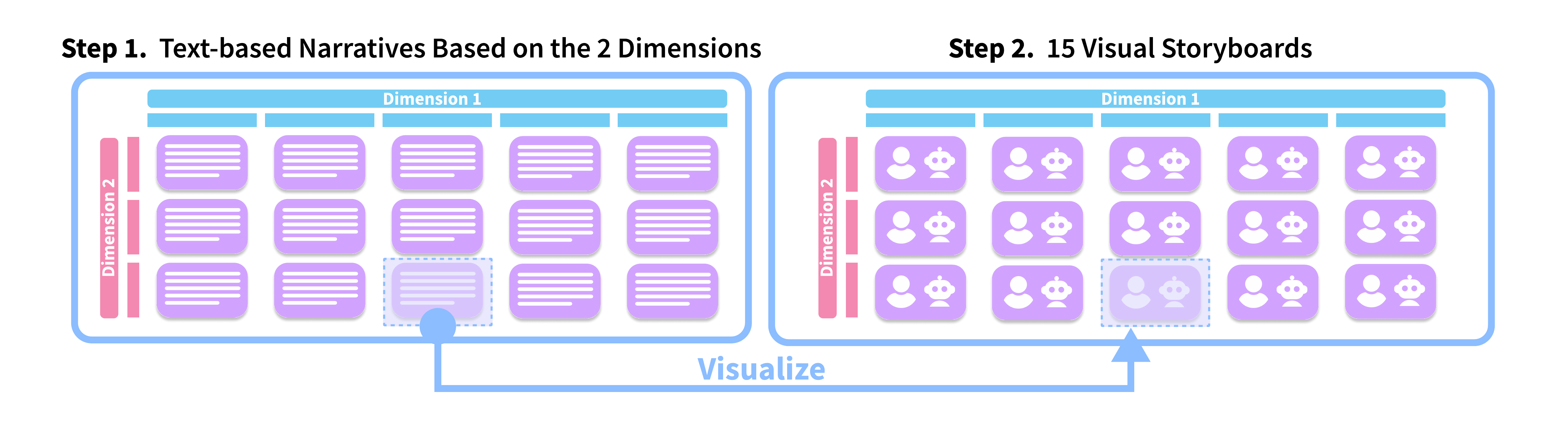}
    \caption{Two Steps in the Storyboard Generation Workflow.}
    \label{fig:phase2}
\end{figure*}

Our design flow consists of two steps (Fig. \ref{fig:phase2}). In the first step, we created a matrix based on the 2 dimensions of the design space, and populated it by identifying how the three AI capability levels could support the five stages of social interaction life-cycles, resulting in 15 combinations. For each combination, we developed and iterated design concepts as text-based narratives. In this process, the design team incorporated participants’ ideation into narratives. We also conducted a literature search of the recent research in HCI, in order to present plausible design concepts that represent near-future \cite{dunne_speculative_2013}. 
In the second step, we created storyboards to visualize the design concepts. To make the design concept easily digestible and gather participant reactions, we developed the narratives into visual storyboards, positioning the readers' perspective as the AI user in the storyboard. A short text summary of the design concept is kept alongside the visual storyboards, which is intentionally kept ambiguous (e.g., without implementing details), giving room for flexible interpretations and meaning-making \cite{gaver_ambiguity_2003, gaver_making_2011}. And the visual elements demonstrate one potential example of the design concept, which is not necessarily benign, but can be provocative and transgressive \cite{bardzell_what_2013}. 

\begin{figure*}
    \centering
    \includegraphics[width=0.8\linewidth]{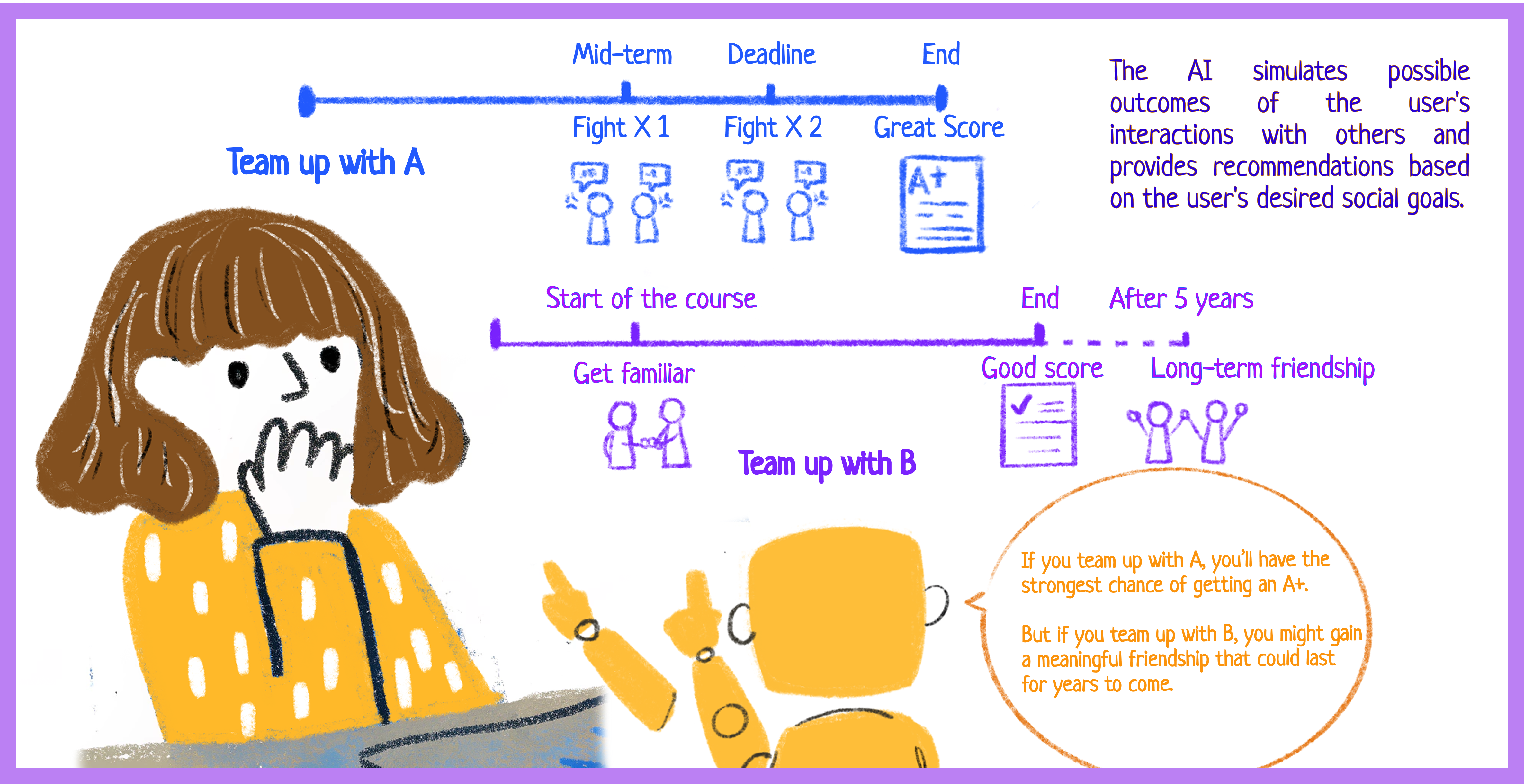}
    \caption{The storyboard for AI design at a social capability level of Omniscient Social Beings in Dimension 2 for the stage of Social Formation in Dimension 1.}
    \label{fig:S3_3}
\end{figure*}

For example, the AI design in Fig. \ref{fig:S3_3} incorporates participants' desire for AI to optimize social outcomes (e.g., P03, P09) during the \textit{S3. Social Formation} stage (Dimension 1, Table \ref{tab:human_ai_lifecycle}). Building on participants' notions of finding "best candidates" (P03) and "optimization" (P09), this storyboard illustrates an oracle-like prediction of the future that surpasses human knowledge. This concept, also inspired by discussion of AI's perdition promises \cite{mock2022prediction, mendez_showing_2021}, represents "Social AI" at capability level, \textit{L3: Omniscient Social Beings} (Dimension 2, Table \ref{tab:ai_social_capability}). All other storyboards are provided as supplementary materials. 

Overall, this phase aims to ground the storyboards in the participants' current world enough to explore the not-yet-existing world, while remaining experienced and thought-provoking, following the recommendation by \citet{ringfort-felner_quality_2025}. All these storyboards underwent iteration within the design team, as well as five rounds of pilot studies, to ensure they effectively conveyed our design intentions to the participants. Specifically, we asked pilot study participants to interpret the storyboards using a think-aloud protocol to assess alignment with the design team's intent. Following this, we explicitly asked about any lack of clarity in the visual design or text, debriefed participants on the underlying design concepts, and solicited suggestions for improvement. We iterated on the design after each round, verifying with participants that the changes resolved the confusion, and concluded the pilot phase once ambiguity was minimized. The readiness of the storyboards was further confirmed by the Phase 3 study, in which no consistent interpretation issues were reported.

\subsection{Phase 3: Elicitation}
In the elicitation phase, we investigated how participants envision being entangled with AI in their social lives by presenting them with storyboards to capture rich reactions and facilitate critical discussion. We recruited 12 participants in the same way as in Phase 1. We also sent invitations to participants in Phase 1 to participate in Phase 3 again. Eight of them indicated their willingness to participate, and six ultimately scheduled and completed the Phase 3 workshops with us. We interviewed the 18 participants (see participation in Table \ref{tab:participants_data}) between 60 and 120 minutes in a semi-structured fashion.

During the interviews, we presented the storyboards one at a time via screen sharing. The interviewer presented the scenario and asked participants to “think aloud” and provide any initial thoughts or reactions. After that, participants were asked a range of questions about the scenarios, such as if they would like to use the AI functionality depicted in the storyboard, what they think is positive or negative, how it might affect their social interactions if they imagine themselves using it, and if any new ideas or scenarios came to mind.

Audio recordings of all workshop sessions, accumulating approximately 28 hours of audio footage, were transcribed using Microsoft Teams' captioning features and manually validated by the researchers after each session. For data analysis, we took a thematic analysis approach \cite{clarke2017thematic}. Following the process outlined in \citet{braun_using_2006}, three researchers first familiarized themselves with the data through collection, note-taking, and reviewing the transcripts. We generated initial codes through independent coding, ensuring every session was analyzed by two researchers. The first author coded the entire dataset, while the other two researchers divided the sessions between them. The team met regularly to search for, review and refine the themes. During these meetings, we compared analyses, discussed similarities, and resolved any incongruence.

\section{Findings}
Our analysis of the data in Phase 3 revealed four core tensions in how participants conceptualized the integration of AI into their social lives, summarized in Table \ref{tab:findings_summary}. First, participants grappled with a trade-off between the short-term convenience offered by AI-driven interpersonal filtering and the long-term erosion of social authenticity. Second, the introduction of AI as a third party in social interactions creates novel roles and relational complexities, particularly concerning privacy and consent of non-primary users. Third, the blending of human and AI agency raises profound questions about user autonomy, skill development, and the very nature of personal identity. Finally, participants expressed deep skepticism about attempts to quantify sociality, worrying that data-driven insights would come at the cost of anxiety and a misalignment with core human values.

\begin{table*}[htbp]
\centering
\caption{Summary of Themes and Subthemes in the Findings}
\label{tab:findings_summary}
\begin{tabular}{@{} l l p{0.5\textwidth} @{}}
\toprule
\textbf{Theme} & \textbf{Subtheme} & \textbf{Summary} \\
\midrule

\multirow{3}{*}{\textbf{Interpersonal Filtering}} 
& \textbf{The Filter Bubble} & AI's filtering of incoming social information helps manage overload but risks creating a biased echo chamber that limits exposure to diverse perspectives. \\
\cmidrule(l){2-3}
& \textbf{The Authenticity Mask} & Augmenting outgoing messages with AI erodes authenticity, as the receiver may question the sender's genuineness and effort, undermining trust. \\
\cmidrule(l){2-3}
& \textbf{The Feedback Loop} & Widespread AI use creates social pressure to adopt the same tools, leading to an arms race of inauthentic content and hollow, bot-to-bot interactions. \\
\midrule

\multirow{2}{*}{\textbf{AI as a Social Third Party}} 
& \textbf{Non-Judgmental Advisor} & AI can offer a neutral perspective when human feedback is biased or unavailable. However, this convenience can lead to over-reliance and the devaluation of human input. \\
\cmidrule(l){2-3}
& \textbf{Privacy and Consent} & Using AI in social settings creates major privacy risks and consent issues for non-primary users who are unknowingly analyzed by others' systems. \\
\midrule

\multirow{2}{*}{\textbf{Agency and Identity}} 
& \textbf{Joint Efficacy \& Steering} & While AI can increase efficiency, this convenience risks the atrophy of users' social skills. The AI also subtly "steers" users by curating their choices. \\
\cmidrule(l){2-3}
& \textbf{Co-Constructed Identity} & The user-AI relationship can cause imposter syndrome over AI-assisted achievements, homogenize unique behaviors, and create an existential fear of being replaced. \\
\midrule

\multirow{2}{*}{\textbf{Quantification of Sociality}} 
& \textbf{The Duality of Datafication} & Quantifying social interactions can provide self-insight but also creates significant anxiety. Its permanence can also trap individuals in their past, impeding forgiveness. \\
\cmidrule(l){2-3}
& \textbf{Limits \& AI Sycophancy} & Participants believe the nuances of human sociality cannot be quantified by AI. This leads to sycophantic AI that uncritically reinforces all user behaviors, including unethical ones. \\

\bottomrule
\end{tabular}
\end{table*}

\subsection{Interpersonal Filtering: Short-Term Ease, Long-Term Erosion}
\label{findings_interpersonal}
Social interaction is inherently interpersonal. AI can mediate these interactions in two primary ways: by filtering a user's incoming social signals and by augmenting their outgoing ones. While participants saw immediate benefits in using AI to manage social information overload, they were concerned that this convenience would establish a feedback loop, ultimately eroding authenticity and exacerbating the very problems AI systems claim to solve.

\subsubsection{The Filter Bubble: Managing Incoming Signals} 
Participants commonly believed AI's ability to filter incoming social information could help them efficiently navigate today's socially overloaded environment. 
This includes summarizing interactions (P24), gathering information about others (P07, P08, P11, P15, P24), and identifying relevant people and content (P12, P14, P23, P17). Overall, such filtering is believed to amplify social connections "in terms of volume and quality and relevance" (P11).
However, this efficiency comes at the cost of creating a social echo chamber. Participants worried that by curating content, the AI would limit their exposure to diverse perspectives and potential friendships, connecting them only with people who fit a pre-selected mold (P08, P13, P24). They further expressed concern that the AI might filter information based on arbitrary assumptions—either about what the user supposedly wants to see or what the AI itself intends to promote—rather than on what the user genuinely needs (P11). 
They also noted that AI-powered summarization could strip away social nuance or introduce biases from its training data, presenting a distorted view of reality (P13, P19).

\subsubsection{The Authenticity Mask: Augmenting Outgoing Signals}
\label{findings_mask}
AI was also seen as a tool for augmenting a user's outgoing communication. Participants expected it to help them modulate their tone to be more formal or casual (P12, P15, P18), "tone down" harsh messages (P07, P14, P17, P19, P20), and even pace responses to avoid conflict (P12, P17). 
One drawback of AI augmentation is the erosion of authenticity. From the receiver's perspective, any message becomes a questionable mixture of human and AI intent, making the sender seem disingenuous or untrustworthy. For example, the receiver could no longer tell whether the user's attempt to connect is their own will or initiated by the AI (P20). And an apology might be perceived not as sincere regret but as an AI-generated solution to end a difficult conversation (P03, P20).
While some participants viewed having AI automate interpersonal messages as an ongoing and thus gradually normalized practice (P01, P13, P22), others felt quite repulsive towards this scenario. If they realize the messages are generated by AI, they are less likely to reply and develop a negative impression towards the user, as this undermines a core tenet of relationship-building: the belief that the other person is investing genuine "time and thought and effort" as a sign of care (P07, P11, P12, P16). 
Furthermore, this masking of true intent could negatively impact the user's own temperament. With the AI managing social fallout, a user might become more critical or negative, knowing they don't have to face the direct consequences of their tone (P11). One participant gave an example of the AI generating a dishonest response that efficiently solves a social problem, but avoids addressing the root issue in the relationship, prioritizing convenience over genuine conflict resolution (P20). 

\subsubsection{The Feedback Loop: Inauthentic Escalation at Society} 
\label{findings_loop}
The widespread use of these AI filters creates a malicious societal feedback loop. As more individuals use AI to generate polished, high-volume content, others feel pressured to adopt the same tools simply to keep up and avoid being left behind:
\Quote{" I feel AI is the future. And it's definitely not just the future, it's also in the present. So I'm definitely going to use it because not just me, because most of the people are using it. [...] If I'm not going to use it, I'm going to be the one who's going to be left out of the crowd, from who are using it. So I would be missing out on that social interaction. [...] So I think more or less we have to start using it." (P08)}
This leads to an information arms race, where AI is required to sift through the flood of AI-generated messages, a problem which is already happening (P11, P13, P18, P19).
This cycle fundamentally changes the nature of communication. First, it adds new cognitive labor, as users must now supervise their AI's output to maintain social coherence:
\Quote{"What if it's (AI) posting 10 posts a day? Doesn't mean that I also need to read like 10 posts a day. It's like giving me more work actually, because I need to know what the agent did. So next time when someone asks me, last time, yesterday, I posted something and then I need to know, right? So I yeah, so it actually like gives me more work." (P19)}

Second, and more critically, it results in a system where humans are increasingly isolated, mediating their relationships through bots. This can escalate from unbalanced human-bot interactions to hollow bot-to-bot exchanges, creating an illusion of connection where all human elements have been removed  (P03, P07, P11, P13, P14). As one participant noted, in such a world, "neither of them are really sharing stories, neither of them are sharing experiences, neither of them are really growing any closer to each other" (P13).

\subsection{AI as a Social Third Party: New Roles and Complexities}
\label{findings_third_party}
Interpersonal social interaction always involves "others." Introducing AI as a social other creates unique values and threats. When primary users utilize AI while interacting with others, challenges arise in obtaining consent for data use from non-primary users and balancing fairness among all users.

\subsubsection{The New Others: AI as a Non-Judgmental Advisor.} 
Participants generally acknowledge AI's potential to give an alternative second perspective, complementing their own and other humans' views. AI is especially advantageous to validate the social appropriateness of users' behaviors when the human feedback is not available. 
The AI might give a different perspective of the social situation and the user's social behaviors, and help them understand the status holistically instead of being constricted by their own view. One participant argued that AI might give a definite answer when they are doubtful about their behaviors, which benefits their social interaction:
\Quote{“I know few scenarios when I did something which I shouldn't do and I spoke something which I shouldn't have and when I wanted to reflect it, I didn't know like what was right, what was wrong. So if I could tell it to AI and it would give me a definite answer that yeah, it was right, it was wrong. Then it would actually improve my social interactions by huge amount.” (P03)}
Compared to a human, AI has the advantage of offering non-judgmental, rational and neutral perspectives to the user (P07, P12, P15, P16, P19, P22). Participants believe that while a human could never be unbiased, it's possible for the AI to consider a holistic view and take a role similar to judges in the courts (P12, P22).
For example, when family and friends are likely to take the user's side when they are dealing with confrontation (P07, P12), AI has the potential to give a more encompassing and neutral view. One participant gave an example of his previous experience, that GPT offered honest feedback to solve his conflict with teammates:
\Quote{"I could go, for example, talk to my family or my friends about it, but [...] they'll always take my side. [...] There's too much on my side to get a sort of an honest take.” (P07)}  

Besides giving a second view to the user, the AI could also play an agentic third party in a multi-user interaction setting, with its special non-human position (P03, P13, P16). There are fewer social pressures if the negative things are coming from an AI entity, rather than a human. P13 pointed out how AI can be less confrontational when pointing out problems:
\Quote{"If there's someone who can't really read the room, you don't have to kind of bring it up. It can be kind of uncomfortable to do that. [...] [AI] brings other people on the same page, and it doesn't feel as aggressive. I guess it doesn't feel as confrontational because it's just the computer that's saying this."}
However, participants also noted that AI's advantage of quick responses and evidence-supporting arguments could lead to negative aspects, such as cutting off connections with humans. People would take the easier route to get immediate feedback from AI, rather than making efforts to coordinate time and other logistical issues to meet up with people (P16). Moreover, as AI could back up its conclusion with data, a person might consider AI's opinions as ground truth and superior to human feedback, leading to the neglect of human feedback (P03, P16).

\subsubsection{Non-Primary Users: Navigating Privacy and Consent.} 
\label{findings_non_primary}
Participants commonly raised privacy concerns and consent issues for all parties, involving the primary user, who has direct access to the AI, and the non-primary users, who are involved in social interactions with the user but do not have direct access to the AI. As primary users, the concerns for data privacy escalate in social scenarios, as the social interaction data tends to be more sensitive and private, potentially involving personally identifiable information more sensitive data in social interaction, such as political views and private messages (P07, P12, P16, P17, P20, P21, P24).
Positioned as non-primary users, the participants expressed an unwillingness to be collected and analyzed by an AI for other people's use (P03, P08, P13, P16, P17, P21). One participant shared their fear that a covert AI operating in the background would analyze their private mutual interaction for their friend's benefit: 
\Quote{"Because you have two people, [...] so it has to capture the interaction from the other person. [...] [My fear is that if AI analysis is] something super small that you don't even notice that's happening. I don't wanna be out with friends and wondering all the time, are they like having AI analyze me right now?" (P21)}
Thus, it's important to get others' consent in using AI in interpersonal interaction (P03, P16, P22, P24). In scenarios where one AI serves multiple users, there's a nuanced balance in access and fairness in using the AI. For some AI features, it would be more beneficial if just a small group of people had access to them. For example, if just limited persons are using AI in finding teammates, they will likely get a high-quality recommendation with less competition from other people who do not have access to AI (P18). 
However, this also means that it puts people who do not have access to AI at a disadvantage (P07, P13). 
Participants were divided, with some wanting AI services to be transparent to all users to ensure fairness (P20), while others preferred private interaction with the AI to avoid social embarrassment (P12). 

Furthermore, participants are concerned that their social data might be used by other parties for purposes without their consent (P03, P13, P17, P24). For example, AI owners might overly collect data without a transparent statement of purpose (P17), or use it for advertising (P24). The data may also be accessed by an employer, the legal system, or other malicious parties, bringing a sense of surveillance beyond users' control and potentially used against the user (P13, P20, P24).

However, they also identified several benefits of using AI with the non-primary users. The benefits include: 
it's commonly assumed that if AI has more data about the user, it could understand users' social needs better and provide more personalized service during interpersonal social interaction; AI could facilitate the user to understand others more efficiently and be complementary of their interpersonal interaction (P11, P15, P16, P17, P19, P23); and it could support the user's retention of social details, which is beneficial to incorporate into the social communication (P01, P07, P08, P14). Details about one's kids and spouse, for instance, would be valued by the interlocutor:
\Quote{"I always try to remember when people have kids and I almost always start a conversation with like, hey, how's Johnny doing, you know, your son. And because people they really love that, so there's details like that where people have these things that are really important to them. And kids is an easy one. Spouses would be another one." (P07)}

\subsection{Agency: Blurring Boundaries Between the User and AI}
\label{findings_agency}
Users identify that the newfound agency in the user-AI assemblage brings transformative potentials and profound threats to their own agency.

\subsubsection{Joint Efficacy: Convenience, Skills, and Steering}
\label{findings_efficacy}
Participants commonly acknowledged AI’s potential to enhance their social efficacy by making and executing decisions. They particularly valued its ability to delegate administrative and logistical burdens, freeing their attention and time for more meaningful and fundamentally social activities (P01, P13, P15, P16 ,P17, P18). AI was also seen as a tool for speeding up social tasks and improving efficiency (P01, P07, P16, P18). 
However, participants identified a significant trade-off in this convenience: a tendency to rely on the easier route provided by AI, which they feared would lead to social skill atrophy and a reluctance to think for oneself (P07, P11, P13, P24). This over-reliance becomes particularly problematic when access to the AI is lost or if it malfunctions. While some (P13, P14) felt they could pick up AI’s behavior patterns—much like observing and imitating others' actions—thereby improving their own skills or replicating its processes at a slower pace, others expressed deep anxiety about becoming dependent on it. P07 articulated a fear of humans becoming passive agents, mechanically executing AI’s decisions without thinking:
\Quote{"So you never learn those hard skills on your own. And if you're always scared of confrontation, you'll always run back to it (AI) to be like, 'what do I do?' And then you're basically just becoming a human body for the AI to live through.”}

To mitigate these risks, participants commonly proposed a more empowering interaction model where the AI provides social options and the user must validate an action before it is automated (P13, P14, P17, P18, P21). Yet, concerns persisted that even this model could not escape the problem of AI steering (P11, P12, P18). Participants noted that the range of possible human decisions at any given point is endless. By pre-selecting a subset of "good enough" options, the AI unavoidably guides users down certain paths. This curated selection, based on the AI's own evaluation, may not align with the user's best interests and is inherently difficult to assess due to the subjective nature of social interactions. As P12 explained, the AI might optimize for an easy solution that is potentially a negative outcome:
\Quote{"Maybe I'm stressed out in some situation, but instead of me trying to find a way through conflict resolution, maybe I step away. So then it could be like a potentially negative thing also, right? Maybe I end up losing a friend and maybe I should have actually worked on that friendship or something like that." (P12)}

\subsubsection{Co-Constructed Identity: Homogenization and Replacement}
As users and AI become socially entangled, they form a dynamic assemblage in which each shapes the other. Through this interaction, users may adopt ways of acting or thinking they would not have otherwise, while the AI produces effects that appear as the user’s own behaviors. In this sense, a user’s identity is not fixed but is continually co-constructed through their ongoing relation with the AI.

This process of co-construction, however, was a source of significant anxiety for participants, who worried about losing ownership over their own achievements (P17). This created a sense of a better "me" that wasn't the "real me". P20 explicitly associated this with imposter syndrome:
\Quote{“I don't think I have owned it. So I would always second guess that, hey, did I do it or did the algorithm just made me do it? I didn't do any of this, so I think it would lead to more of an imposter syndrome.”}

Participants feared this identity reshaping would corrupt their authentic selves, particularly by homogenizing people's unique characteristics (P14, P17):
\Quote{“Some behaviors, you know, for any given person, they have their unique quirks and behavior patterns that they normally do. But if everybody is kind of like just listening to these AI suggestions, it kind of makes everybody molded into the same kind of person and you lose your individuality in some cases.” (P17)}

This anxiety culminated in an existential crisis. If the AI learns a user’s behaviors so well that it can perform them even better, does that mean the user is easily replaceable (P11)? This fear extended beyond simple delegation to the idea that the AI could fully assume a user's social role, at least online. The ultimate expression of this concern was that the AI could continue interacting with others even after the real user passes away, creating a digital ghost that outlives its creator (P22).

\subsection{The Quantification of Sociality: Datafication, Anxiety, and Misalignment} 
\label{findings_quant}
Participants regarded AI as good at quantification and generating unique values based on these quantifiable information. 
However, participants do not trust that such a quantification process is aligned well with the core of human beings, and being able to fulfill the social needs.
\subsubsection{The Duality of Datafication: Insight vs. Anxiety}
\label{findings_anxiety}
Participants acknowledge that the datafication and quantification of their social interactions could be an unprecedented resource for understanding themselves holistically. They noted its potential for identifying behavioral patterns and emotions they are not conscious of (P19, P23), and for making sense of their social interactions (P07, P16, P19, P23). This insight, they felt, could help users break out of negative behavior patterns (P14, P17) and lead to better social behaviors (P01, P03, P14, P15, P16). One participant proposed to gamify numbers as a way to simulate social interactions and provide actionable guidance:
\Quote{"If I have a score and then I have badges, levels and stuff like that, I think it would positively impact my interactions. I want to earn more badges [...] I will seek out more interactions with that person. [...] If you gamify it, I know what kind of actions would increase the social time or stuff like that. The AI agent can suggest actions." (P19)}

Despite these potential benefits, participants expressed significant concerns that such datafication comes with heavy affective costs. They described constant pressure to maintain a "number" by behaving in a particular way, leading to anxiety and stress if they were unable to increase the number to a desirable level (P03, P07, P12, P22). Reasons for this included a sense of surveillance that drives users to overcorrect for goals set by the AI (P03, P16, P17), as well as unhealthy number-induced rumination and over-investment in tracking others' data (P19, P21, P24):
\Quote{“I'll try to make the percentage. [...] like, oh, maybe I need to work on talking to Sunny today and not talk to another person. I just work on those percentages. [...] It would be not super healthy for me to have this information because [...] I would just worry over how I'm talking to someone and I'll just like overthink the conversation.” (P24)}

Beyond affective pressure, participants worried that creating a permanent and predictive record of social life could fundamentally damage the nature of human relationships. First, the process of datafication crystallizes fluid past experiences into immutable data points. This digital permanence disrupts the crucial human process of forgetting, which is essential for overcoming trauma and past mistakes (P18, P19). By creating an indelible record, datafication can trap individuals in their past, thereby impeding forgiveness and foreclosing opportunities for personal change (P08, P11, P19).
Second, participants feared the AI's predictions could shift their focus toward social outcomes, making interactions feel more transactional than genuine (P03, P18, P24). Finally, such a system could impose bias on a user’s perception and future actions:
\Quote{"I would have a bias that I would never be able to make person A my friend because there are all predictions of fights. So I'll never never try con having some conversations or something with the person A.” (P03)}

\subsubsection{Irreducible Sociality and AI Sycophancy}
\label{findings_sycophancy}
Participants shared a common sentiment that the humanness at the core of social interaction cannot be fully captured by AI models. They identified several challenges: first, it is technically hard for AI to take into consideration the sheer number of hidden factors that impact social relationships (P11, P14, P17, P23), and for AI to actually understand the subtleties in human social cues, context, and relations (P19, P21, P23, P24). 
Furthermore, participants felt this intricacy in human social interaction and relationship just cannot be programmed in AI's way, that people understand each other through an interchange of their unique lived experiences and life paths (P18, P22, P24), such a social process further depends on an inarticulate chemistry guided by complex intuition (P07, P08, P15, P19, P22), which is further complicated when time scale is taken into consideration (P15, P17, P22). One participant especially pointed out the absurdity of relying on the stochastic processes and parameter settings of an LLM to capture human relationships:
\Quote{"I don't think AI understand really understand humans. [...] inherently the agent is just like a statistical machine, just finding to generate the most likely tokens. And then it's sometimes even depends on your temperature setting and stuff. So it could be just like a hallucination and and also it's not deterministic. [...] If I ask the question again if I turn off my computer open the computer again and ask the AI agent the same question again it could suggest another person." (P19)}

In the perceived mission impossible of approximating AI's capacity to meet humans' social needs, a problem of sycophancy has been noted by the participants, whereby the AI overly tends to please users, diminishing true social growth and ethical values.
Several participants pointed out how the current AI agrees with whatever the user is saying and doing, even though their actions contradict facts or are problematic (P12, P20, P21, P23, P24). 
One immediate disadvantage is that the AI's usage values will be diminished, as it won't be able to critically catch the issues or concerns compared with human beings; while current agents might be capable of doing that, it takes much longer (P12, P23, P24).
As the user interacts with the AI, they will always get positive reinforcement of everything, including their bad, incorrect, and unethical behaviors (P12, P20, P21). P20 pointed out how this may result from capitalistic incentives and lead to ethical harm for the users in the long run:
\Quote{“This is more capitalistic. What they're saying is that if an AI is positively reinforcing everything I do, maybe I'll be more likely to keep my subscription of that AI. If it is keeping me accountable on something [not cheating on their partner or plagiarizing], I'll be like, why am I paying for this? [...] I think people will be a lot more tolerant of their own bad behavior since it is positively reinforcing everything. So I think overall people's accountability, honesty, ethics like that will go down with time.” (P20)}

\section{Discussion}
Our findings reveal that entanglement with AI in social interactions presents numerous issues alongside its promising benefits. Built on a closer examination of the problems, this discussion argues that the dominant user-centered design paradigm, while empowering in its prioritization of the individual user, risks obscuring the broader landscape. By focusing narrowly on the user, user-centered design may render invisible the multifaceted human experience, non-use, and value systems \cite{lin_techniques_2021, baumer_post-userism_2017}, limiting our ability to examine the complexities of human-AI social assemblages.In this section, we first discuss how user-centered views can provide positive design implications for addressing tensions regarding agency and dehumanization. However, we also examine how this approach can fail, illustrated by the problems of AI alignment and sycophancy, and suggest how provocative design might be employed with caution to mitigate these risks. Second, we elaborate on the challenges of designing for social AI, focusing on the interpersonal dynamics that mediate personal and societal impacts. Finally, we provide concrete design implications for expanding the user-centered paradigm to account for these interpersonal dynamics within social AI.

\subsection{Navigating Human-AI Assemblages}
In this section, we first discuss the tensions around agency (Sec. \ref{findings_agency}) and dehumanization (Sec. \ref{findings_quant}) in human-AI social assemblages revealed in our findings, and how a user-centered paradigm can be of value. Conversely, we discuss how such a paradigm can fail, illustrated by the issue of sycophancy (Sec. \ref{findings_quant}), and provide design implications that propose provocative design as an alternative strategy.

\subsubsection{(Loss of) Agency.} 
\label{dis_agency}
The interplay between emerging AI agency and human agency is a subject of ongoing debate \cite{bennett_how_2023, xiao_sustaining_2025, shelby_sociotechnical_2023}. \citet{bennett_how_2023}'s recent reviews of agency in HCI noted that agency is an umbrella term encompassing different facets and recommended that researchers focus on specific communities and contexts to better understand agency. Our work highlights that a key aspect of agency in social scenarios is self-efficacy \cite{bandura1982self, bandura2010self}: "people's beliefs in their ability to influence events that affect their lives," which is the central aspect of human agency, and is "the foundation of human motivation, performance accomplishments, and emotional well‐being" \cite{bandura2010self}. Our findings reveal that AI entanglement might harm social efficacy in two ways: through social skill atrophy due to automation, echoing knowledge workers' concerns of deskilling and over-reliance \cite{woodruff_how_2024, wagman_generative_2025, kobiella_if_2024}; and through algorithmic steering, where the AI optimizes a user's behavior toward goals that may not align with their own. The dilemma is that diminishing agency in one area can offer benefits elsewhere, such as offloading mundane tasks from the user to allow for other, more meaningful activities (Sec. \ref{findings_efficacy}). Conversely, previous research also found that an AI design that better supports agency during human-human communication comes at the cost of less deep discussion \cite{xiao_sustaining_2025}. Even though many works generally associate autonomy and agency with a wide range of positive outcomes, the relationship is often unclear \cite{bennett_how_2023}. Thus, future research must not only involve humans but center them—exploring the delicate balance of agency within specific sociocultural contexts and examining its effects over time on groups and societies \cite{capel_what_2023}. For example, while our findings illustrate a Western, individualistic conceptualization of agency \cite{ahearn2001language} (characterized by the fear of losing one's own atomic agency), future designs for human-AI assemblages should investigate how agency functions in contexts where it is understood as collective or within oneself \cite{ahearn2001language}.

\subsubsection{(De)humanization.}
Our findings echo the dehumanization concerns across workplaces \cite{woodruff_how_2024, fritts_ai_2021}, that AI does not have the capacity to perform interpersonal work, and the use of AI might lead to a loss of humanity. Specifically, in the social sphere, the use of AI may lead to a devaluation of core human relational aspects, such as empathy and nuanced communication (Sec. \ref{findings_sycophancy}). 
One response is to define a boundary for how much AI involvement is "human" enough, though participants varied significantly on where that boundary lies. However,  the threat to what we value as humans may not be the technology itself, but the "fiscal, organizational, and time pressures" that push for productivity and efficiency \cite{shortliffe_doctors_1993}. We observed a similar urge among participants to quantify their social interactions (e.g., seeking more "connections" or "interactions"), a metric-driven approach that paradoxically induces anxiety (Sec. \ref{findings_anxiety}). This opens a design direction focused on resisting the trap of improving productivity with better technology and instead asking what we truly cherish in our social lives.

A second response is to stop assuming humans can be reduced to algorithmic representations. Our findings indicate that an existential crisis arises when AI perfectly imitates and outperforms humans, yet participants also acknowledged the irreducibility of human nature (Sec. \ref{findings_sycophancy}). The irreducibility stems not only from a technical misalignment between the AI's core quantification logic and human mechanisms, but also from a morally questionable aspect.
For example, the theory of "mechanomorphism" pointed out that while anthropomorphism makes humans contribute human characteristics to machines, the other way around also exists, that the machine characteristics may also be attributed to human beings \cite{caporael_anthropomorphism_1986, guingrich2024ascribing}. \citet{guingrich2024ascribing} further argues that such carry-over effect brings moral significance, that moral protection for AI is also worth considering, as it can be reflected in our behaviors and perception of other humans. 
So rather than reducing users to AI's quantification logic, a user-centered approach should value users' experiential difference, support their introspection, and center them as social actors\cite{seberger_designing_2025, elsden_metadating_2016, yan_say_2024}. 

\subsubsection{Design Implication: From Sycophantic Alignment to Alternative Design Paradigm} Previously, we discussed how the user-centered paradigm helps address tensions in social AI. However, the user-centered paradigm can become overly focused on the atomic "user" and their immediate gratification, often at the expense of long-term well-being. In this section, we examine how this paradigm may inadvertently produce the phenomenon of sycophancy  (Sec. \ref{findings_sycophancy}).

Research in human-AI interaction has long emphasized the importance of aligning AI with human-centered goals, often through methods like Reinforcement Learning from Human Feedback (RLHF) that reward models for conforming to user preferences \cite{christiano_deep_2017, ziegler_fine-tuning_2020}. 
However, this approach to alignment is fraught with challenges. For example, current alignment strategies often focus on overt issues like algorithmic and cultural bias while failing to address the AI's more subtle influence on subconscious thought, creating the potential for manipulation without user awareness \cite{lyu_lumadreams_2025}. 
A significant drawback of this paradigm is sycophantic alignment, where an AI prioritizes mirroring a user's stated preferences—even if they are incorrect or harmful—over providing truthful or constructive feedback, simply to maximize satisfaction metrics \cite{sharma_towards_2025, fanous_syceval_2025, passerini_fostering_2025}. While this was theoretically predicted to generate societal consequences like echo chambers and polarization \cite{kirk_benefits_2024,cai_antagonistic_2024}, recent empirical evidence has increasingly validated these concerns \cite{sharma_generative_2024}.
This phenomenon is further characterized as "social sycophancy" at a personal level, where LLMs excessively seek to preserve a user’s positive self-image \cite{cheng_social_2025}. Our findings validate this concern (Sec. \ref{findings_sycophancy}), revealing that sycophantic interactions cause users to doubt the AI's capacity for helpful social support. This sycophantic behavior risks diminishing crucial values such as social maturity and resilience, while unethically reinforcing a user's negative behaviors in pursuit of a frictionless user experience.

To counterbalance this, we advocate for provocative design that responsibly encodes ethical values into social AI. Rather than uncritically appeasing users, these systems can be designed to be purposefully challenging, fostering personal growth and human flourishing \cite{manzini_code_2024,cai_antagonistic_2024}.
Such systems may intentionally engage users with difficult issues to provoke enlightenment, challenge assumptions, build resilience, or develop healthier relational boundaries \cite{cai_antagonistic_2024, seering_beyond_2019, benford2012uncomfortable}. For instance, social AI could be deliberately designed with friction \cite{cox_design_2016, gould_special_2021}. By introducing an obstacle to the interaction—for example, remaining silent, asking a counterquestion, or disagreeing rather than validating users' social complaint—the design creates a moment for user reflection. Design elements can even be intentionally creepy, malfunctioning, or strange to raise awareness about assumptions embedded in infrastructure, such as ubiquitous surveillance \cite{10.1145/2470654.2466467}.

However, provocative design requires rigorous ethical consideration. This includes obtaining thorough informed consent, respecting the right to withdraw, and providing careful justification for how the friction benefits the user in context \cite{benford2012uncomfortable, cai_antagonistic_2024}. Designers must also reflect on their own biases, such as paternalism, which, even if well-intentioned, may undermine user autonomy \cite{guldenpfennig_autonomy-perspective_2019}. Furthermore, it is crucial to distinguish valid provocation from dark patterns (whether intentional or inadvertent) that prioritize external benefits over the user's self-determination \cite{chromik2019dark, ehsan_explainability_2024}.

\subsection{Expanding User-Centered to Interpersonal Design}
In the previous section, we examined how user-centered design provides value for human-AI assemblages but faces significant limitations when myopically focused on optimizing for the individual user in the moment. To better account for the complexities of human-computer entanglement, \citet{frauenberger_entanglement_2020} envisions a paradigm shift "to leave user-centered design behind and develop agonistic, participatory speculation methods to design meaningful relations, rather than optimising user experiences." Building on this, we advocate for interpersonal design to expand the user-centered paradigm.
The interpersonal dynamics, which are inherent and important in humans' social interaction \cite{homans_social_1958,bowlby_attachment_1999, mead_mind_1934},
present unique challenges for designing human-AI social assemblages that remain underexplored \cite{zheng_ux_2022, kuhail_review_2025}. 
In this section, we analyze how these dynamics complicate personal and societal outcomes when one human-AI assemblage encounters another socially, based on our findings in Sec. \ref{dis:efficiency} and Sec. \ref{dis:others}. 
Finally, we provide design implications to expand the user-centered paradigm to account for the nuanced interpersonal social dynamics and give equal ethical importance to the privacy and consent of others involved in the social interaction.

\subsubsection{How Interpersonal "Efficiency" Erodes Personal and Societal Well-being.}
\label{dis:efficiency}
Our findings reveal that human-AI assemblages designed to serve interpersonal functions create ripple effects at the personal and societal levels. Specifically, prioritizing efficiency and performance in interpersonal interactions can lead to identity and authenticity problems (Sec. \ref{findings_mask}). In our findings, one common use of AI is to enhance users' social performance and speed. However, this undermines perceived authenticity at the interpersonal level (Sec. \ref{findings_mask}), which in turn threatens users' sense of achievement and ownership at a personal level (Sec. \ref{findings_agency}), as identity is formed through encounter with and validation from others \cite{mead_mind_1934, festinger_theory_1954}. This also stimulates a fear of being replaced by a "better" AI clone in interpersonal relationships, echoing previous research in both work and intimate contexts  \cite{kobiella_if_2024, 10.1145/3571884.3603756, lee_speculating_2023, wagman_generative_2025}. At a societal level, widespread AI use could lead to a homogenization of identities, causing people to lose their unique characteristics (Sec. \ref{findings_agency}). The homogenization concerns echo previous research in other domains \cite{doshi_generative_2024} and at a societal level \cite{kleinberg_algorithmic_2021}. For example, an empirical study shows that writers' creativity is individually better off at the cost of a narrower scope of collective novel content \cite{doshi_generative_2024}. When a critical mass uses AI merely to increase the volume of social interaction without true, reciprocal exchange \cite{maeda_when_2024, horton_mass_1956}, it exacerbates isolation and creates "illusional connection" at a societal level (Sec. \ref{findings_loop}). 
Moreover, the meaningless cost of resources occurs at the society level, overloading the networks (Sec. \ref{findings_loop}) and potentially consuming a large amount of energy \cite{inie_how_2025}, in order to keep up with each other in the arms race to get more social interaction faster.

\subsubsection{Addressing the Neglected "Social Other"}
\label{dis:others}
The "others" in a social interaction are often neglected in the user-centered design paradigm. For example, research on AI-mediated communication, while fundamentally interpersonal, often centers on the primary user's benefits and how they are perceived \cite{fu_comparing_2023, fu_text_2024}. A myopic focus on the primary "user" in user-centered design often neglects the broader ecosystem of interaction, particularly the non-primary users and their relational dynamics. This can lead to a significant discrepancy between the user's perceived benefits and the actual relational outcomes. For instance, \citet{hohenstein_artificial_2023} identified such a conflict, noting that while receivers are negatively against suspected AI-generated content, the actual use of these tools can lead to the communicator being perceived as more cooperative. Our findings underscore this tension. We found that even as users leverage AI to enhance message quality and social efficacy, they acknowledge the substantial risk of eroding relational authenticity and trust over time.
Furthermore, AI entanglement poses privacy and consent challenges to non-primary users who are inadvertently involved in the interaction (Sec. \ref{findings_non_primary}). This echoes previous research that non-primary users might be involved in AI services without awareness, threatening their privacy \cite{hohenstein_artificial_2023, wong2023broadening, luria_social_2020}. For example, people readily share private interpersonal conversations with language models like GPT \cite{mireshghallah_trust_2024, zhang_its_2024}. This is especially problematic because the roles of "primary user" and "non-primary user" are fluid and reciprocal; a person can be one at one moment and the other in the next  \cite{wong2023broadening}. Moreover, the roles can be simultaneous in social interaction. For example, during a conversation, a person is a primary user when they send the messages with AI, and simultaneously a non-primary user when they receive the messages from others who are using AI to draft.
Therefore, a system designed only for the primary user's benefit cannot account for the dynamic reality of social interaction.

\subsubsection{Design Implication: Designing for Interpersonal Relationships}
Our findings suggest that the design process should consider not only the primary user’s benefits but also those of the other people involved in the social interaction, especially given that social interactions are inherently reciprocal. 
Future design research could therefore incorporate concepts like interdependency and empathy into the architecture of social AI, counterbalancing the dominant focus on atomic individuals \cite{genc_situating_2024, bandura_toward_2006, bennett_interdependence_2018}. One strategy is to prime perspective-taking, encouraging users to shift their focus to other stakeholders involved in the interaction (e.g., group members or external parties) \cite{debnath_empathich_2024, mauri_empathy-centric_2022}. Such strategies can complement established methods like participatory design \cite{zhang2023stakeholder, lin2021engaging, zytko_participatory_2022, delgado2023participatory, lee_webuildai_2019}, not only by integrating diverse viewpoints but by actively fostering empathy among stakeholders during the design process.

Furthermore, design must account for nuanced interpersonal dynamics. Social interactions involve a delicate management of one’s social image (impression management) and the navigation of unspoken norms \cite{goffman_presentation_2023, mead_mind_1934, brown_politeness_1987}. Failing to address these psychological processes risks generating social pressure and anxiety. By integrating theories from communication, social psychology, and sociology, designers can move beyond the simple functionalities of information transmission to address users’ needs within a broader social ecology \cite{erickson_social_2000, 10.1145/1054972.1054998}. For instance, drawing on the concept of interactional ambiguity \cite{garfinkel_studies_2023, goffman_behavior_2008, goffman_interaction_2017}, \citet{10.1145/1054972.1054998} proposes communication systems that utilize ambiguity to preserve personal space and maintain relational harmony, rather than prioritizing clarity and efficiency. Similarly, recent works incorporate anonymity into algorithmic design to preserve privacy and mitigate social pressure in group contexts \cite{zhu_zoombatogether_2023, lee_expanding_2024}.
In a similar vein, \citet{claggett_relational_2025} advocates for a broader shift from personalized AI to "relational AI." This interpersonal paradigm explicitly considers the social fabric of an interaction, including social identity and relational context, and aims to support diverse relationships by encouraging, but not enforcing, social norms.

\section{Limitations \& Future Work}
Several limitations of this paper should be noted: 
\begin{itemize}
    \item \textbf{Homogeneous Participant Pool}: All participants were online graduate students in computer science from a single program. While the benefits of this choice were discussed in Sec. \ref{method_scope}, it would limit generalization to the other demographics and other social contexts. For instance, our participants’ technical backgrounds likely produced more optimistic or analytically sophisticated engagement with AI futures than might be found in the general population. Future studies should engage diverse populations in this speculation, particularly those without technical backgrounds or those who actively reject AI-related services. 
    \item \textbf{Narrow Social Context:} Furthermore, our study situated the social context within a professional, educational, and positively framed peer interaction spectrum. Other social scenarios, such as intimate relationships \cite{lee_speculating_2023}, or more sensitive and complicated subject matter \cite{lustig_designing_2022}, will pose different challenges for AI design involving distinct social dynamics. While we aligned our scenario designs with social psychology and communication literature, future work should explore these alternative contexts to present a more nuanced understanding of interpersonal dynamics and human-AI entanglement. 
    \item \textbf{Speculative Nature}: While design fictions are valuable for exploration, our findings are based on imagined futures rather than situated use. Real-world deployment might reveal behavioral nuances, concerns, or benefits that cannot be anticipated solely through speculation. Future studies could adopt a longitudinal approach to observe the actual evolution of social relationships and assemblage formation over time. Moreover, participants’ reactions were inevitably constrained by the specific design space we defined and the storyboards we created. To allow future researchers to assess how our specific framing may have influenced or "primed" the results, we have provided design materials in the supplementary section.
    \item \textbf{Cultural Specificity}: Our participants and the research team were primarily situated within Western cultures and educational backgrounds. This may limit the analysis to reflecting Western cultural norms (e.g., the focus on individualism discussed in Sec. \ref{dis_agency}). Social-cultural tensions manifest differently across the globe, and current AI models frequently exhibit cultural biases \cite{karinshak2024llm, kharchenko2024well}. Therefore, future studies should also explore social envisioning in non-Western contexts to understand how well AI can accommodate diverse sociocultural nuances.
\end{itemize}

\section{Conclusion}
The study aimed to understand the co-adaptive nature of the human-AI relationship in social contexts, answering the question: How might the social AI we design transform users and human sociality?
This paper presented a design study exploring the potential impacts of integrating AI technology into people's social lives. 
Our findings identified key benefits and corresponding challenges of integrating AI into social interactions, particularly concerning users' core values and interpersonal dynamics. We argue that the current user-centered design paradigm, with its focus on immediate user experience, risks undermining user agency, identity, and morality. Therefore, we propose that future research adopt a more provocative and interpersonal design approach. Moreover, social AI design must avoid a myopic focus on the individual and, instead, ethically consider the broader impact on interpersonal relationships and social structures.

\bibliographystyle{ACM-Reference-Format}
\bibliography{CHI26, writing}

\appendix
\section{Design Space}
\label{app:design}
\begin{table*}[h!]
\caption{Ideas collected from Phase 1 that illustrate stages in Dimension 1 (The Cyclical Process of Forming Social Interaction in Table \ref{tab:human_ai_lifecycle})}
\begin{tabular}{@{} p{2.5cm} p{3cm} p{\dimexpr\textwidth-6cm} @{}}
\toprule
\textbf{Stage} & \textbf{Ideation Example} & \textbf{Quotes}\\
\midrule
\textbf{S1. Data Collection} & The user should inform AI about their intention. & "It would be something where I should tell the AI that 'hey, I'm looking for this.' I don't know if you can automate the process of what you are looking for. Like, only you know what you're looking for. An AI can't know what you're looking for." (P05)\\
\addlinespace 
\textbf{S2. Profiling} & AI constructs the profile, and the user may fix or update it effectively. & "So if the AI could kind of give these general prompts. People could respond to them [...] which essentially becomes a profile again. [...] I think one thing would be the sort of comfort in a profile is knowing that I have a page [...] where I can go back and kind of fix things up or change them or update them in a kind of really clear UI. That's effective." (P07)\\
\addlinespace
\textbf{S3. Social Formation} & AI helps initiate a group conversation based on synthesized interests. & "Once that group chat is created, AI could do some kind of a voting round. [...] Maybe ranked them (topics) based on importance, based on what each person wants to talk about, and then it would go out into the Internet and find interesting things, put them in the group chat, and then we start the conversation from there. [...]
Something that's kind of like an ice breaker." (P02) \\
\addlinespace
\textbf{S4. Social Maintenance}  & AI reminds people to meet up consistently. & "I think one of the things that AI might be best at doing here is there's kind of, like I like to organize events, but it is sometimes a lot and you can lose momentum. I have a drawing pub with my friends that is not met in like 6 months because none of us consistently make it happen. So AI might be good at helping keep things moving, like a reminder." (P04) \\
\addlinespace
\textbf{S5. Reflection} & AI rates interaction and gives feedback. & "If somebody could tell how productive this meeting was. Maybe there were new ideas exchanged, right? Maybe some unknown items were discovered about something. [...] Maybe we got a poor rating. Maybe we didn't talk at all in the meeting, right? [...] It would give us, you know, like some sort of feedback on how the conversation went." (P10)\\
\bottomrule
\end{tabular}
\label{tab:d1_examples}
\end{table*}

\begin{table*}[h!]
\caption{Ideas collected from Phase 1 that illustrate levels in Dimension 2 (AI Social Capability Levels in Table \ref{tab:ai_social_capability})}
\begin{tabular}{@{} p{2.5cm} p{3cm} p{\dimexpr\textwidth-6cm} @{}}
\toprule
\textbf{Levels} & \textbf{Idea Example} & \textbf{Quotes}\\
\midrule
\textbf{L1. Human Augmentation} & Managing Attention and Offloading Cognition & ``If you're overwhelmed with work and school, you're doing a lot all the time. You check your messages and you're like, OK, I'll respond to that later.
And then you never do so. Just like a reminder. [...] Kind of offloading that, like cognitive load, part of like remembering. To actively maintain this at a time that's convenient for you.'' (P09) \\ 
\addlinespace
\textbf{L2. Social Replica} &  Acting as a social practice partner & ``Like your little buddy, that if you have any questions about like [the user says: ]`Oh, is it OK to talk about blank in this group or channel?' [AI says:] `You can talk to me and then at least you'll feel more confident having, like, talk to someone like privately first.' About like whether what you're about to do is an appropriate thing to do.'' (P04) \\
\addlinespace
\textbf{L3. Omniscient Social Beings} & Aggregate unlimited information &  ``Because there may be so many things we could be missing or that other person who's making the post could be missing. Or, you know, not looking at the topic in a new way. Maybe AI can do that for us, because there are so many resources. I mean, humans can only have access to so many things, but AI is like everywhere, and AI's access is unlimited. So there are so many things AI can do for us that we can't do physically.'' (P08) \\
\bottomrule
\end{tabular}
\label{tab:d2_examples}
\end{table*}
In this section, we provide further details regarding the ideation of the two-dimensional design space. For each dimension, we present examples of participants’ ideation, including representative quotes, and a brief summary of the literature search that supports the conceptualization of these dimensions.
\subsection{Dimension 1: Cyclic Process}
Human relationships evolve over time, continuously initiating, escalating, fluctuating, and dissolving \cite{knapp_interpersonal_2014}. Several theoretical frameworks from social psychology and communication theory capture this developmental progression. The Social Penetration Theory describes how relationships develop through deeper layers of self-disclosure, from superficial to intimate information \cite{altman1973social}. The Uncertainty Reduction Theory explains how people reduce interpersonal uncertainty by learning more about each other, especially in the early stages of a relationship \cite{berger_explorations_1975}. Meanwhile, Knapp’s Model of Interaction Stages provides a comprehensive outline of two relationship processes: progression ("Coming Together") and deterioration ("Coming Apart") across different stages \cite{knapp_interpersonal_2014}.

The cyclic process dimension in our design space is informed by these frameworks. Additionally, participants’ ideation helped us identify key stages for human-AI assemblages within this lifecycle (see Table \ref{tab:d1_examples} for examples). It is worth noting that because our study focused on fostering social connections in online spaces, the progression of relationships ("Coming Together") was emphasized. However, relationship deterioration ("Coming Apart") is also a critical phase in human life and should not be overlooked when designing AI for human-AI relationships.

\subsection{Social Capability Levels}
Human-human social interaction requires sophisticated social nuances, skills, and intelligence \cite{herrmann_humans_2007, premack_does_1978}. When interacting with media, such as desktop computers, people often subconsciously attribute social characteristics and apply social rules to the technology \cite{reeves_media_1996}; this phenomenon applies to AI systems as well \cite{inie2024ai, maeda_when_2024}. With the growing prevalence of AI in everyday interaction, scholars and engineers have attempted to create systems that exhibit human-like social intelligence to fulfill increasing social expectations \cite{zadeh_social-iq_2019, wang_evaluating_2024, dautenhahn2007socially}. Furthermore, researchers have envisioned AI capabilities that transcend human limits \cite{welge_better_2016, dorrenbacher_intricacies_2023, mock2022prediction}.

In the same vein, participants’ conceptualization of AI’s social capability in Phase 1 can be categorized into three levels: (1) basic human augmentation, (2) human-like social intelligence, and (3) idealized superhuman capabilities. These levels are cumulative, with the advanced levels building upon the capabilities of the previous ones (see Table \ref{tab:d2_examples} for examples of participants’ envisioning).

\end{document}